# Crime Analysis using Open Source Information

Sarwat Nizamani[1,*], Nasrullah Memon[2,3], Azhar Ali Shah[4], Sehrish Nizamani[5], Saad Nizamani[5], Imdad Ali Ismaili[4]

[1]Department of Computer Science, Sindh University Campus Mirpurkhas
[2]The Maersk McKinney Moller Institute, University of Southern Denmark
[3]Mehran University of Engineering and Technology
[4]Institute of Information and Communication Technology, University of Sindh
[5]Department of Information Technology, Sindh University Campus Mirpurkhas

[*]email: sarwat@usindh.edu.pk

**Abstract-**In this paper, we present a method of crime analysis from open source information. We employed un-supervised methods of data mining to explore the facts regarding the crimes of an area of interest. The analysis is based on well known clustering and association techniques. The results show that the proposed method of crime analysis is efficient and gives a broad picture of the crimes of an area to analyst without much effort. The analysis is evaluated using manual

## 1. INTRODUCTION

Crime analysis is one among the essential component of any law enforcement agencies. Crime analysis can help police agencies in reducing the crimes (Bruce et al., 2008). Crime analyst's task includes the review of crime reports on daily basis. The method presented in this paper will help crime analysts to understand the law and order situations of an area of interest, which utilizes the open source information. Open source information also referred as open source intelligence is defined by Best (Best, 2008) as the "retrieval, extraction and analysis of publicly available sources.". The major source of open source information in the current era is the Internet, which makes available the information from diverse sources. The roots of the term open source intelligence are affiliated with the security services as well as to the law enforcement agencies (Best, 2008). Open source information is an asset for many of the stakeholders. The information is available in enormous volume, which can be processed in different ways to gain knowledge.

Open source intelligence can be regarded as a kind of unclassified information, which intentionally focuses ad hoc audience to address particular questions (Steele, 2007). Open source information can be collected from various sources, Internet being the major source of information. The Internet sources include the news channels, social media sites, government and non-government organizations' press releases, blogs, forums and others. The process which involves open source information follows certain steps.

These steps include retrieving required information from various sources, pre-processing the information, analyzing the pre-processed information according to the required task and finally draw the conclusions.

The process which is carried out in this study is depicted in **(Fig. 1),** which clearly illustrates the process, first the information is gathered from divers sources (Internet being the main source). Next the required piece of information is extracted and necessary pre-processing is performed. Afterwards, the information is analyzed in order to highlight the major issues of the area using the proposed method i.e. clustering association. Finally, conclusions are drawn from the analysis and the evaluation.

This paper is an extension of author's previous study of semantic analysis of FBI (Federal Bureau of Investigation) news reports (Nizamani and Memon, 2012). In the study (Nizamani and Memon, 2012), a semantic analysis of FBI news reports is carried out, which returns the important information of the news by highlighting the facts present in each news report. In the current study, we do not analyze the news reports individually; the analysis is performed on the bulk of the news/ press releases, which reveals important facts regarding crimes in the specific area. The analysis helps in understanding the current law and order situations in the area, so that specific actions may be taken by law enforcement agencies in order to deal such situations. The accessibility of the cyber-space has favored the news channels and other national and international organizations in facilitating to make available the news in enormous quantity. Analyzing the news manually is an irksome task, in order to deduce conclusions. News are generally found as un-structured or semi-structured text documents. Data mining techniques are often used for discovering the concealed knowledge from the available sources of information.

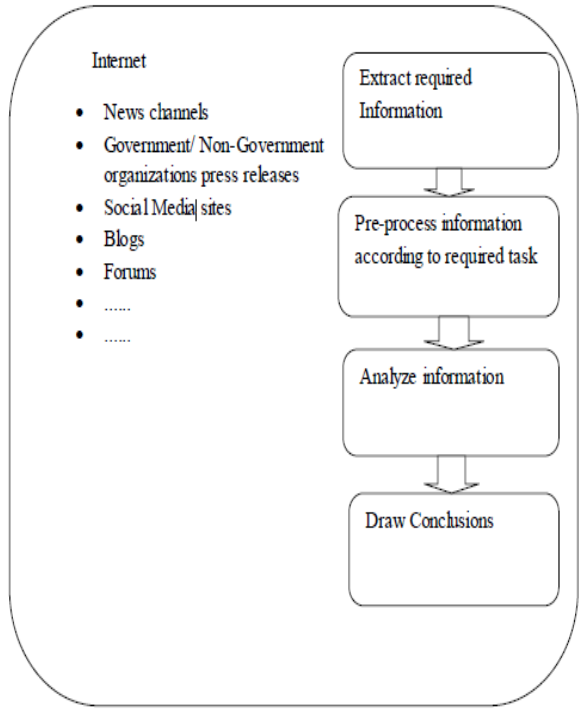

**Fig.1: Open source information analysis process**

Chen *et al.* (Chen *et al.*, 2003), in a study presented an overview of crime data mining using case studies. The study employed the police narrative reports for mining crime related information. Simultaneously, the authors mentioned that obtaining these reports is very difficult. The study presented in this article is aimed to exploit open source information i.e., FBI press releases, in order to get insight of the news of an area. In the FBI press releases most of the reports are more or less about the crimes and criminals, because FBI[1] is United State agency for crime investigation. The method of analysis proposed in this paper gives a better understanding of the law and order situations of an area and other crime related matters. The purpose of the study is to analyze the crimes in a generic way, but not at individual level. The proposed method of analysis highlights the common crimes in the area and also details the convictions of the criminals. The analysis caters a broad picture of the overall crimes in an area of interest. Rest of the paper is structured as follows: Section 2 presents related works; while Section 3 discusses the crime analysis model. Finally, Section 4 concludes the paper along with future work.

## 2. RELATED WORK

[1] http://www.fbi.gov/news/stories

Crime analysis has been performed from different perspectives and in diverse dimensions by the academicians. It has remained the area of interest for researcher of different fields such as, sociologists, criminologists, computer scientists, physicists, mathematicians and others. In the last few decades, there has been greater development in computer and the Internet related technology. The advancement in computer technology has contributed to crime analysis task to greater extent. With the use of computer-aided technology for crime analysis, the crime rate is dropped in US (Sengupta *et al.*, 2014). The computational models have great potential for analyzing crime related activities (Brantingham *et al.*, 2009).

Crime analysis can vary from task to task and the open source information can be utilized in a number of ways depending on the type of analysis. For instance, criminal network analysis is an essential type of crime analysis, which analyzes the globally organized crimes.

The study (Jennifer and Chen., 2005) proposed a framework for the investigation of criminal networks, which is named as crimeNet Explorer. There are four major phases of the framework, including, (a) network creation; (b) network partitioning; (c) structural analysis; and (d) network visualization. Chen *et al.* (Chen *et al.*, 2003), in the article discussed four case studies using data mining techniques in the project COPLINK. These include: (a) entity extraction from police narrative reports; (b) detection of criminal identity deception; (c) authorship identification in cybercrimes; and (d) criminal network analysis.

The advancement of the Internet sources has privileged the news channels to the tremendous amount of news. It would just be the waste of cyber-space, if the volume of information is not utilized properly (Nizamani, 2014).

Despite these computational models, in the literature there exist a number of mathematical models for analysis of criminal activities. For instance, the articles (Lindelauf *et al.*, 2008; Maeno & Ohsawa, 2009; Memon *et al.*, 2007) exploited graph theory for the study of covert networks (Fellman *et al.*, 2015; McGahan, 2015 ).

Mathematical approach of game theory (Sandler *et al.*, 2003) is also widely used by researchers for crime analysis. The criminals and police are considered to be the players, while analyzing crimes using game theory.

Researchers from the field of Physics have also contributed in the field of crime analysis. Galam, (2002) modeled the role of passive supporter in global threat of terrorism using the concept of percolation. The author emphasizes that in the current situation of terrorism the entire world is under the terrorism risk and the military operations are insufficient in this regard.

Thus, almost all of the studies discussed above, use the open source information for crime analysis. In this paper we use open source information in the form of FBI press releases, which help in understanding the crimes of an area.

## 3. MATERIAL AND METHODS

In this paper, we analyzed crimes of an area of interest by exploiting the well-known un-supervised techniques from the field of data mining. On this basis, the paper proposes a cluster and association based analysis method, which gives a finer intuition of the law and order conditions of certain area. This elongate and elucidative process of the crime analysis based on clustering and associations is illustrated in **(Fig. 2)**. All the steps of the crime analysis model are implemented using R data mining and statistical tool (Zhao, 2012). The analysis begins by retrieving heading sentences from a collection of FBI press releases. The next step is to pre-process the headlines for further analysis and statistical information extraction. The statistical analysis includes the frequencies of the terms found in the news heading (it should be noted that simple term frequencies are extracted). In order to extract term frequencies *TermDocumentMatrix*[2] function is used from *tm* (text mining package of R) package. For association well known method *Apriori* algorithm is used. Once the statistical analysis is carried out, then the terms of the news headings are visualized using word cloud[3], which provides better apprehension of the major news/crimes. Based on the statistical analysis, the clusters of the frequent terms are constructed, which groups the terms in close relationship.

In the paper, for case study, the crime analysis of the Albuquerque[4] city of New Mexico[5] has been carried out. The word cloud of the news is given in **(Fig. 4),** which clearly caters the hint regarding the prominent issues in the area. The visual analysis of the word cloud elucidates salient terms in the news.

Afterwards, the association analysis is carried out which clearly provides the information about major issues. The associations of the prominent terms of the news are given in Tables 1-5. The cluster analysis of the terms returns the associated terms in the common clusters.

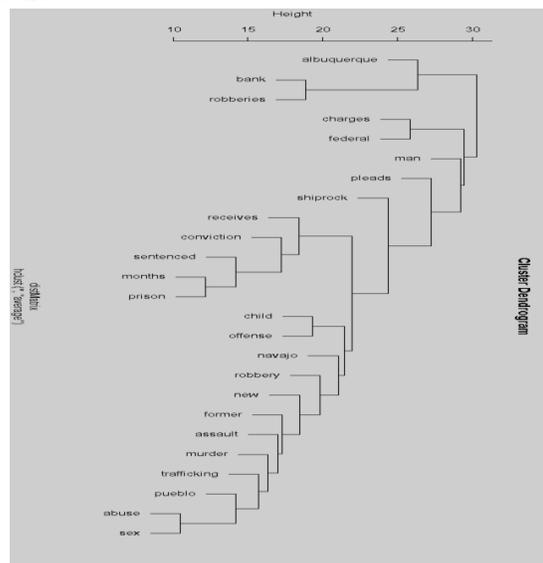

**Fig. 3: Hierarchical Cluster of news terms**

Hierarchical cluster of the terms is depicted in **(Fig. 3)**, which shows that how the terms of news are related? In the next step, the associations of the frequent terms have been determined that give a clue regarding important issues in the area. The associations of the terms, returns the evidence regarding co-occurrences of the terms with a minimum support, which we have set to 0.25. For example, if we are interested in the news regarding ''bank'', the association of the term with other terms will show that what kind of news are about the ''bank''. From the corpus, we see that the news regarding bank are mostly on the bank robberies, because the term ''bank'' and ''robbery'' are correlated.

Once the news reports are visually analyzed, the most prominent terms obtained during visual analysis are then used to acquire the knowledge about co-occurrence relations with other terms found in the news. For instance, in the news reports of Albuquerque city the term "child" is found 61 times out of 326. **(Table 1)** illustrates the associations of the news headings containing the term "child" with other related terms, which reveals the critical facts about "child" related concerns in the area.

In the news headings, another spectacular term found is "bank", which is found 86 times in the corpus. The associations of the term "bank" are extracted with other co-occurred terms, it shows that "bank" frequently occurs with the term "robbery", as shown in **(Table 2).** It discloses that "bank robberies" are one of the major crime in the city of Albuquerque.

---

[2] Each news heading is considered as a document and words in the headings are considered as terms.
[3] Word cloud enhances the visibility of the prominent terms in the text.
[4] Most populated city of New Mexico
[5] New Mexico state is chosen because of enough crime data available in the dataset

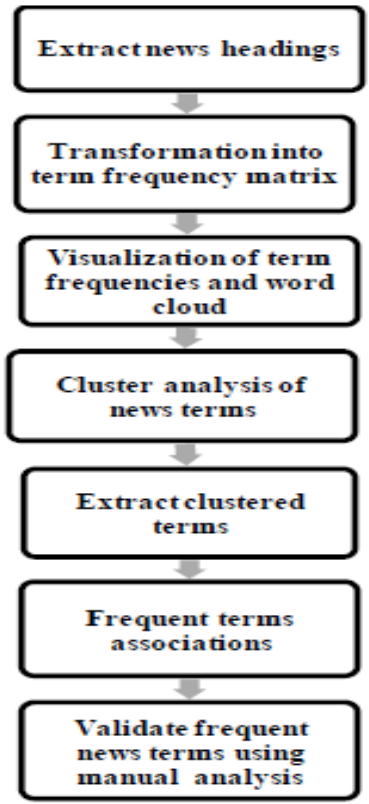

**Fig.2: Cluster and association based news analysis process (Nizamani, 2014)**

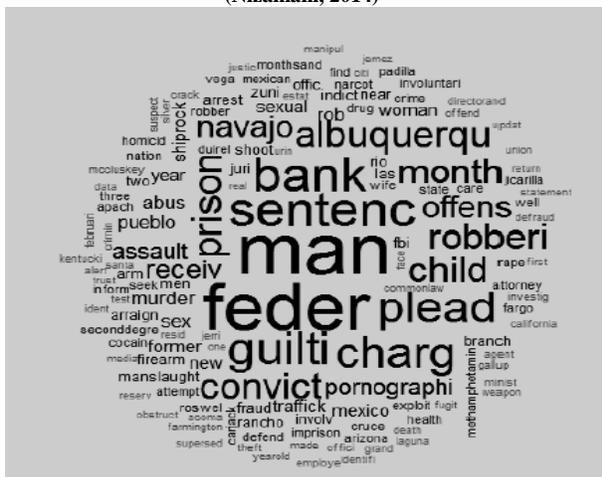

**Fig.4: Word cloud extracted from news of Albuquerque, New Mexico (Nizamani, 2014)**

**Table 1: The co-occurrence of term "child" with other terms in news (Nizamani, 2014)**

| Term | Co-occurrence. |
| --- | --- |
| sex | 0.52 |
| abuse | 0.48 |
| offense | 0.42 |
| sexual | 0.35 |
| exploitation | 0.29 |
| man | 0.27 |
| sentenced | 0.25 |

**Table 2: The Co-Occurrence Of Term "Bank" With Other Terms In news (Nizamani, 2014)**

| Term | Co-occurrence |
| --- | --- |
| robberies | 0.71 |
| robbery | 0.45 |
| albuquerque | 0.43 |

In the news of Albuquerque city, a number of news are about the "sentencing of man in prison". Table 3 illustrates the relationship of "sentenced" to other terms. It can be determined in the word cloud that the term "man" is the most striking term which appeared 144 times in the news headings. When we further analyzed the term using associations, it gives clear understanding of the term's involvement in the news. **(Table 4)** delivers the co-occurrence relationship of the term "man" with the other terms in the news.

Another frequent term found in the news reports is "assault" which appears 31 times in the news reports. When it is further analyzed using associations, it is observed that the term mainly is associated with a tribe "navajo" of United States. **(Table 5)** demonstrates the co-occurrence of term with other terms in the news. The un-supervised analysis of the news unveils the major news of an area.

In order to evaluate the un-supervised news analysis, the manual analysis of the news headings is carried out. The manual analysis is performed by manually reading the news reports. The manual analysis reveals that the unsupervised analysis provides enough information about the major news of the area. For example, the term "child" appeared 61 times in the news headings, the manual analysis shows that there were 61 news related to child in the press releases. Thus, the news analyzed by the proposed method provide same amount of information as one can get it by reading the reports. Hence, proposed method of analysis is efficient in a way can get enough information by saving much of the time.

**Table 3: The co-occurrence of the term "sentenced" with other terms in news (Nizamani, 2014)**

| Term | Co-occurrence |
| --- | --- |
| prison | 0.80 |
| months | 0.72 |
| conviction | 0.69 |
| receives | 0.53 |
| year | 0.37 |
| man | 0.33 |
| imprisonment | 0.26 |
| child | 0.25 |

**Table 4: The co-occurrence of the term "man" with other terms in news (Nizamani, 2014)**

| Term | Co-occurrence |
| --- | --- |
| Federal | 0.33 |
| Navajo | 0.33 |
| sentenced | 0.33 |
| Assault | 0.31 |
| Months | 0.29 |
| Pleads | 0.28 |
| Prison | 0.28 |
| Child | 0.27 |

**Table 5: The co-occurrence of the term "assault" with other terms in news (Nizamani, 2014)**

| Term | Co-occurrence |
|---|---|
| **Man** | 0.46 |
| **Navajo** | 0.31 |

## 4. CONCLUSIONS

In this paper, we presented a crime analysis method from open source information using un-supervised data mining techniques. We employed the clustering and frequent pattern mining techniques. The analysis has been evaluated manually, which shows that the proposed crime analysis method is efficient which gives enough information regarding crimes of an area, and is comparable to manual processing. The manual processing involves human efforts and time. The results show that, by using the proposed method of analysis, the crime analyst can have a broad picture of law and order situations of an area of interest. At present the paper uses press releases as case study, while in future, we plan to analyze social media content when there happen to be some issues and people express their attitude towards that issue on social media. The analysis would help the responsible officials to plan the strategies to resolve those issues.

## ACKNOWLEDGMENT

The part of the paper is taken from the Ph.D. thesis (Nizamani, 2014) of the first author.